\documentclass[aps,prl,twocolumn,floatfix,superscriptaddress]{revtex4-1}

\usepackage{rotating}
\usepackage{epstopdf}
\usepackage{graphicx}
\usepackage{natbib}
\usepackage{epstopdf}
\usepackage{amsmath}
\usepackage{amssymb}
\usepackage{mathbbol}
\usepackage{xcolor}
\usepackage{lipsum}
\usepackage{amssymb}
\usepackage{mwe}
\usepackage{lipsum}
\usepackage{siunitx}
\usepackage[normalem]{ulem}

\usepackage{empheq}
\usepackage{afterpage}
\usepackage[colorlinks,citecolor=blue,urlcolor=red]{hyperref}
\usepackage{cleveref}

\newcommand{\beq}{\begin{equation}}
\newcommand{\eeq}{\end{equation} \smallskip}
\newcommand{\beqy}{\begin{eqnarray}}
\newcommand{\eeqy}{\end{eqnarray} \smallskip}

\newcommand{\bit}{\begin{itemize}}
\newcommand{\eit}{\end{itemize}}
\newcommand{\bmat}{\begin{pmatrix}}
\newcommand{\emat}{\end{pmatrix}}

\begin{document}

\title{Onset of vortex clustering and inverse energy cascade in dissipative quantum fluids}

\author{R. Panico}
\affiliation{CNR NANOTEC, Institute of Nanotechnology, Via Monteroni, 73100 Lecce, Italy}
\affiliation{Dipartimento di Matematica e Fisica E.~De Giorgi, Universit\`a del Salento, Campus Ecotekne, via Monteroni, Lecce 73100, Italy}

\author{P. Comaron}
\affiliation{Institute of Physics, Polish Academy of Sciences, Al. Lotników 32/46, 02-668 Warsaw, Poland}
\affiliation{Department of Physics and Astronomy, University College London,
Gower Street, London, WC1E 6BT, United Kingdom}

\author{M. Matuszewski}
\affiliation{Institute of Physics, Polish Academy of Sciences, Al. Lotników 32/46, 02-668 Warsaw, Poland}

\author{A. S. Lanotte}
\affiliation{CNR NANOTEC, Institute of Nanotechnology, Via Monteroni, 73100 Lecce, Italy}
\affiliation{INFN, Sez. Lecce, 73100 Lecce, Italy}

\author{D. Trypogeorgos}
\affiliation{CNR NANOTEC, Institute of Nanotechnology, Via Monteroni, 73100 Lecce, Italy}

\author{G. Gigli}
\affiliation{CNR NANOTEC, Institute of Nanotechnology, Via Monteroni, 73100 Lecce, Italy}
\affiliation{Dipartimento di Matematica e Fisica E.~De Giorgi, Universit\`a del Salento, Campus Ecotekne, via Monteroni, Lecce 73100, Italy}

\author{M. De Giorgi}
\affiliation{CNR NANOTEC, Institute of Nanotechnology, Via Monteroni, 73100 Lecce, Italy}

\author{V. Ardizzone}
\affiliation{CNR NANOTEC, Institute of Nanotechnology, Via Monteroni, 73100 Lecce, Italy}
\affiliation{Dipartimento di Matematica e Fisica E.~De Giorgi, Universit\`a del Salento, Campus Ecotekne, via Monteroni, Lecce 73100, Italy}

\author{D. Sanvitto}
\thanks{Corresponding author: daniele.sanvitto@cnr.it}
\affiliation{CNR NANOTEC, Institute of Nanotechnology, Via Monteroni, 73100 Lecce, Italy}

\author{D. Ballarini}
\affiliation{CNR NANOTEC, Institute of Nanotechnology, Via Monteroni, 73100 Lecce, Italy}
\affiliation{Dipartimento di Matematica e Fisica E.~De Giorgi, Universit\`a del Salento, Campus Ecotekne, via Monteroni, Lecce 73100, Italy}

\begin{abstract}
Turbulent phenomena are among the most striking effects that both classical and quantum fluids can exhibit. While classical turbulence is ubiquitous in nature, the observation of quantum turbulence requires the precise manipulation of quantum fluids such as superfluid helium or atomic Bose-Einstein condensates. In this work we demonstrate the turbulent dynamics of a 2D quantum fluid of exciton-polaritons, hybrid light-matter quasiparticles, both by measuring the kinetic energy spectrum and showing the onset of vortex clustering. We demonstrate that the formation of clusters of quantum vortices is triggered by the increase of the incompressible kinetic energy per vortex, showing the tendency of the vortex-gas towards highly excited configurations despite the dissipative nature of our system. These results lay the basis for the investigations of quantum turbulence in two-dimensional fluids of light.
\end{abstract}

\maketitle

%
The complex dynamics of turbulent flows keeps attracting the interest of scientists across many fields of research~\cite{Frisch1995}. Quantum turbulence, which studies the turbulent motion of quantum fluids, was initially motivated by the experimental observation of superfluid helium and, later, it has been mostly studied in Bose-Einstein condensates (BEC) of ultracold atoms~\cite{Barenghi2014,white2014}. Despite important differences, quantum turbulence shares observables with its classical counterpart and allows a simpler description in terms of vortices with unitary topological charge~\cite{Tsubota2008}.
Recently, the ability to form highly oblate BEC boosted the research on vortex clustering, an elusive feature of two-dimensional quantum turbulence (2DQT) strongly related to the inverse energy cascade observed in classical 2D turbulence~\cite{Bradley2012,Reeves2013prl,Billam2014,simula2014prl,Groszek2018,Tsubota_Onsager,BEreview}. %
Onsager proposed an explanation of the phenomenon of clustering by applying equilibrium statistical mechanics to a model of point-like vortices: in a closed and conservative system, the formation of quantum vortex clusters is the result of a transition to a thermal equilibrium state which comprises more energy but less entropy (resulting in a negative absolute temperature) as compared to a configuration of randomly distributed vortices~\cite{onsager49}. Vortex clustering and negative temperature regimes were realized only recently in an atomic BEC, thanks to the fine tuning of the excitation conditions achieved in state-of-the-art experiments~\cite{johnstone2019evolution,gauthier2019giant}.  While these studies are centered around atomic BEC, which are well-isolated from the environment and conserve the number of particles for the typical experimental duration, here we investigate 2DQT in a new, dissipative system such as a quantum fluid of light~\cite{carusotto2013quantum}. One main advantage of optical systems is the direct access to the phase of the quantum fluid, based on the analogy between quantum gases and nonlinear optics\cite{weitz2021,Arecchi1991,Ballarini2020,Fontaine2018}.

As a paradigmatic family of quantum fluids of light, here we study exciton-polaritons, bosonic quasi-particles which result from the strong interaction between light and matter in semiconductor microcavities with embedded quantum wells~\cite{Amo2009m}. 
The formation of solitons and vortices in polariton condensates and superfluids has been observed under different configurations, including spontaneous formation due to local density flows, nucleation in the wake of an obstacle and directly imprinted via phase mapping~\cite{lagoudakis2008quantized,Sanvitto2011,Nardin2011,Panico2020}.
Notably, polaritons are strictly 2D, being confined in the plane of the cavity, and the potential landscape can be designed at will both by all-optical and lithographic techniques, which may enable strong connections between topological photonics and 2DQT~\cite{Ozawa2019,Pieczarka2021,Alyatkin2020}. However, few works addressed polariton superfluids in the general context of turbulent dynamics. This is because polariton superfluids, differently from atomic BECs, are inherently dissipative, with the polariton lifetime limited in the picoseconds time range by the photon leakage from the microcavity.
Dissipation sets a strict limit on the timescale of vortex dynamics, which is governed largely by the speed of sound in the fluid~\cite{Galantucci2019}.
This raises the question if 2DQT is even observable in this regime and, in the affirmative case, what are the main differences as compared to a conservative system~\cite{caputo2018topological,zamora2021}. Theoretically, the turbulent dynamics of out-of-equilibrium condensates were first studied by Berloff and, more recently, Koniakhin \textit{et al.} simulated the 2DQT energy spectrum of polariton superfluids in the conservative limit of long polariton lifetime~\cite{Koniakhin2020}. Despite these works suggest that turbulent regimes are indeed possible in polariton fluids, no experimental evidence has been reported so far.

Here we show the onset of vortex clustering and the inverse energy cascade in a polariton quantum fluid. To overcome the intrinsic time limit imposed by dissipation, we create a highly energetic initial state by injecting a polariton superfluid against a potential barrier. We demonstrate that the initial kinetic energy provided to the superfluid is crucial to form a dense vortex gas and accelerate the clusterization dynamics, as confirmed as well by numerical simulations of the polariton nonlinear Schroedinger equation.

\paragraph*{System. -}
\begin{figure}
	\centering
    \includegraphics[width=\columnwidth]{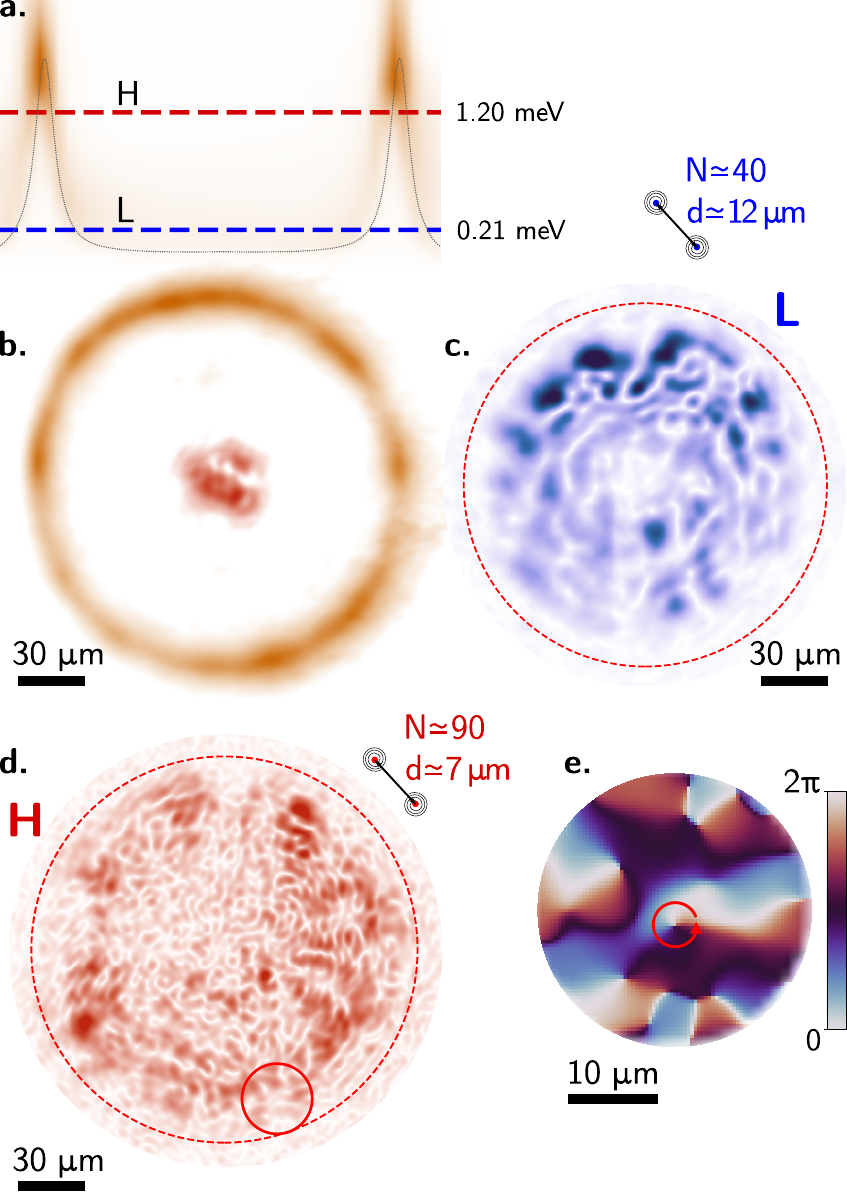}
	\caption{\textbf{a.} Energy resolved photoluminescence of a vertical slice of the ring-shaped potential barrier, created upon non-resonant excitation of the sample. The blue and red dashed lines indicate the injection energies of the superfluid labeled as L and H. \textbf{b.}  Superposition of the image of the trap in real space and the measured density of the polariton superfluid taken 2~\si{\pico\second} after the injection with a pulsed laser. \textbf{c., d.} Time frame of the polariton density taken at 60~\si{\pico\second} after the injection for the low (c) and high (d) detuning, \textit{N} denotes the total number of vortices and \textit{d} the mean distance between nearest neighbouring vortices. The red dashed circle represents the position of the potential barrier. \textbf{e.} Phase of the superfluid corresponding to the red solid circle in (d): vortices can be identified by a $2\pi$ change of the circulation. 
	}
	\label{fig:1}
\end{figure}
In quantum fluids, the formation of clusters of same sign vortices is seen as the statistical signature of an inverse energy cascade. Indeed, clustering not only limits pair-vortex annihilations and the consequent phonon emission, but drives the system to a highly energetic state~\cite{Angheluta2016,Bagnato2020}. To check if similar dynamics can be observed in a dissipative quantum fluid, the system needs to be initialized in a highly excited state. To this aim, we inject polaritons in a high Q-factor ($>10^5$) microcavity~\cite{snoke2013} by using a 2~\si{\pico\second} pulsed laser beam focused at the center of a ring-shaped potential barrier, as shown in Fig.~\ref{fig:1}a-b (further details in the Methods section). The detuning of the pulsed beam from the bottom of the potential barrier corresponds to the initial kinetic energy of the polariton fluid, which expands radially after the injection and reaches the ring barrier after approximately 40~$\si{\pico\second}$~\cite{alyatkin2021,bloch2010}. 
To assess the role of the initial energy on the dynamics of the vortex gas, we consider two injection energies $\delta E$ from the bottom of the potential, namely $\delta E_L= 0.21$~\si{\milli\electronvolt} and $\delta E_H= 1.2$~\si{\milli\electronvolt}, henceforth referred to as L and H, which are indicated in Fig.~\ref{fig:1}a by the blue and the red horizontal dashed lines, respectively.
The temporal evolution of the vortex gas after the collision with the potential barrier is followed using off-axis digital holography~\cite{Donati14926}.
This interferometric technique allows us to obtain a time resolution comparable to the pulse length and enables the direct measurement of both the polariton density and the polariton phase distributions at each time frame with sub-healing length spatial resolution.
A snapshot of the 2D polariton density after 60~\si{\pico\second} from the injection is shown for the L and H configurations in Fig.~\ref{fig:1}c and Fig.~\ref{fig:1}d, respectively. The identification of the vortex positions, including their circulation direction, is realized by searching for an exact $2\pi$ circulation of the phase around each point (Fig.~\ref{fig:1}e).
Since the net angular momentum of our experiment is zero, the system nucleates only vortex-antivortex pairs (dipoles), preserving the neutrality of the charge throughout the whole dynamics.
In the following, we analyse the spatial configuration of the vortex gas and compare its temporal evolution after H and L initial conditions.

\begin{figure}
	\centering
    \includegraphics[width=\columnwidth]{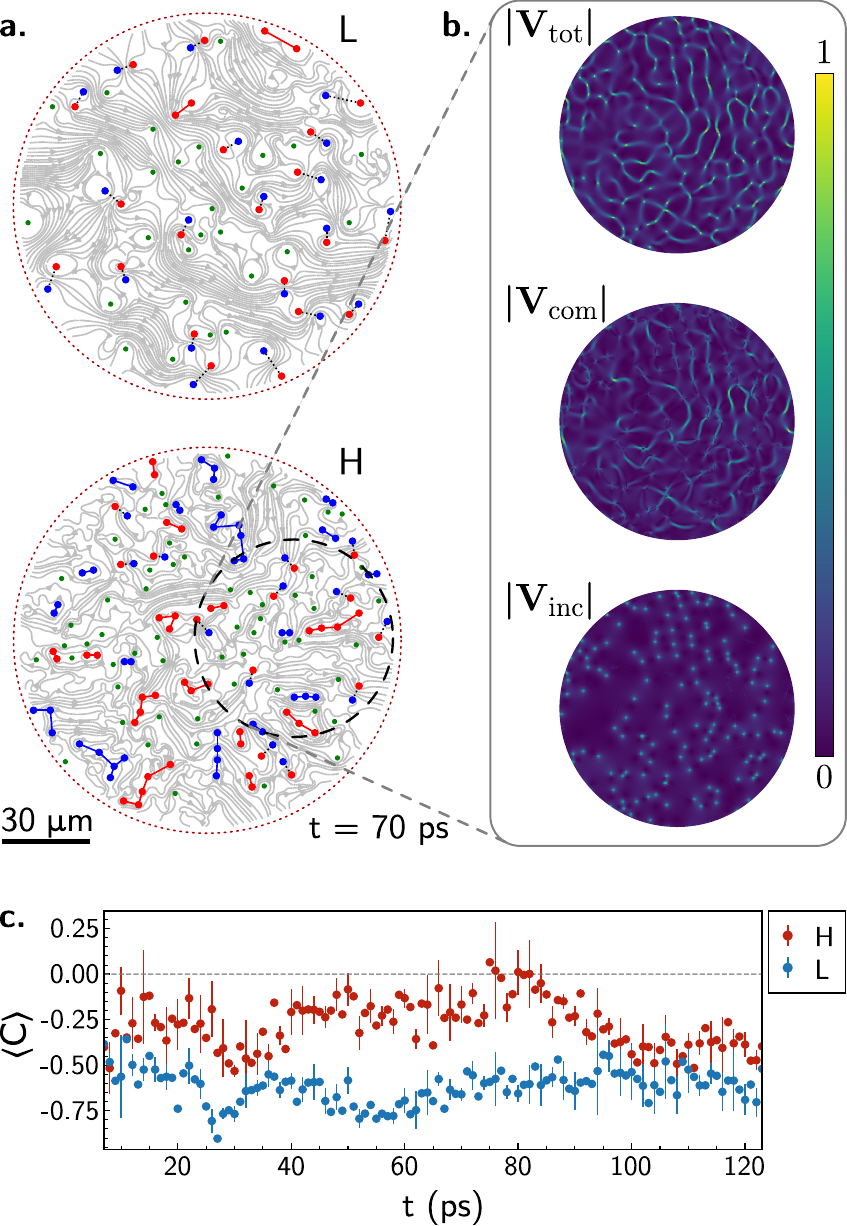}
	\caption{\textbf{a.} Distribution of vortices at low (L, top panel) and high (H, bottom panel) injection energies. Blue and red dots represent positive an negative winding vortices, respectively, belonging to dipoles (black dashed lines) or clusters (solid lines), while for free vortices (green dots) the sign is not reported. The streamlines in the background show the incompressible velocity field of the superfluid. \textbf{b.} Velocity field decomposition of a portion of the superfluid in panel (a). The three figures show the module of the velocity (normalized in each panel), before (top panel) and after the decomposition (compressible and incompressible component in the middle and bottom panel, respectively). \textbf{c.} Vortex first order correlation function for high and low injection energy shown by red and blue points, respectively.
	}
	\label{fig:2}
\end{figure}

\paragraph*{Vortex classification and energy decomposition. -}
In order to study the ordering of our system we classify the phase singularities in three different categories: free vortices, dipole pairs and clusters of the same sign, using the algorithm from~\cite{Reeves2013prl,Valani_2018,johnstone2019evolution}.
Figure~\ref{fig:2}a shows a typical result of this analysis applied to a time frame corresponding to $70$~\si{\pico\second} after the pulsed excitation. The vortex positions are indicated by dots, while the streamlines in the background are the velocity field $\mathbf{v}(\mathbf{r},t) = \frac{\hbar}{m}\nabla\phi(\mathbf{r},t)$, with $m$ the polariton mass, as directly obtained from the measured phase $\phi(\mathbf{r},t)$ of the polariton fluid. In H, the total number of vortices formed after the collision with the border of the ring is larger ($\approx90$) as compared to L ($\approx40$), due to the larger expansion velocity of the fluid. As can be seen from the color code in Fig.~\ref{fig:2}a, vortices mostly belong to dipoles in the upper panel (L), as opposed to the lower panel (H) where the number of vortices in clusters, dipoles and free vortices is comparable. 

Before discussing the temporal evolution, let us stress the importance of being able to easily measure the phase of the superfluid $\phi(\mathbf{r},t)$ at any time interval.
This allows us to separate the contribution to the total kinetic energy due to the presence of quantum vortices from that coming from the sound waves~\cite{nazarenko2014}.
Indeed, applying the Helmholtz decomposition (see \cite{SM}) one can separate the divergence-free (incompressible) and the irrotational (compressible) part of the superfluid velocity, associated with the vortex distribution and the sound waves, respectively. In the next section, we use the incompressible velocity field to compare the information on the kinetic energy with the vortex classification analysis. In Figure~\ref{fig:2}b, the result of the velocity decomposition as extracted from the experimental data are shown for a portion of the fluid in H (dashed circle).
The bright points in the bottom panel of Fig.~\ref{fig:2}b correspond to singularities in the polariton phase. In the tracking and classification analysis, we consider only those vortices with cores separated by a distance larger than the healing length, $\xi = {\hbar}/{(2 m g |{\psi}|^2)^{1/2}}$, with $g$ the interaction constant and $|{\psi}|^2$ the density of polaritons (see Methods). 
\begin{figure}
	\centering
    \includegraphics[width=\columnwidth]{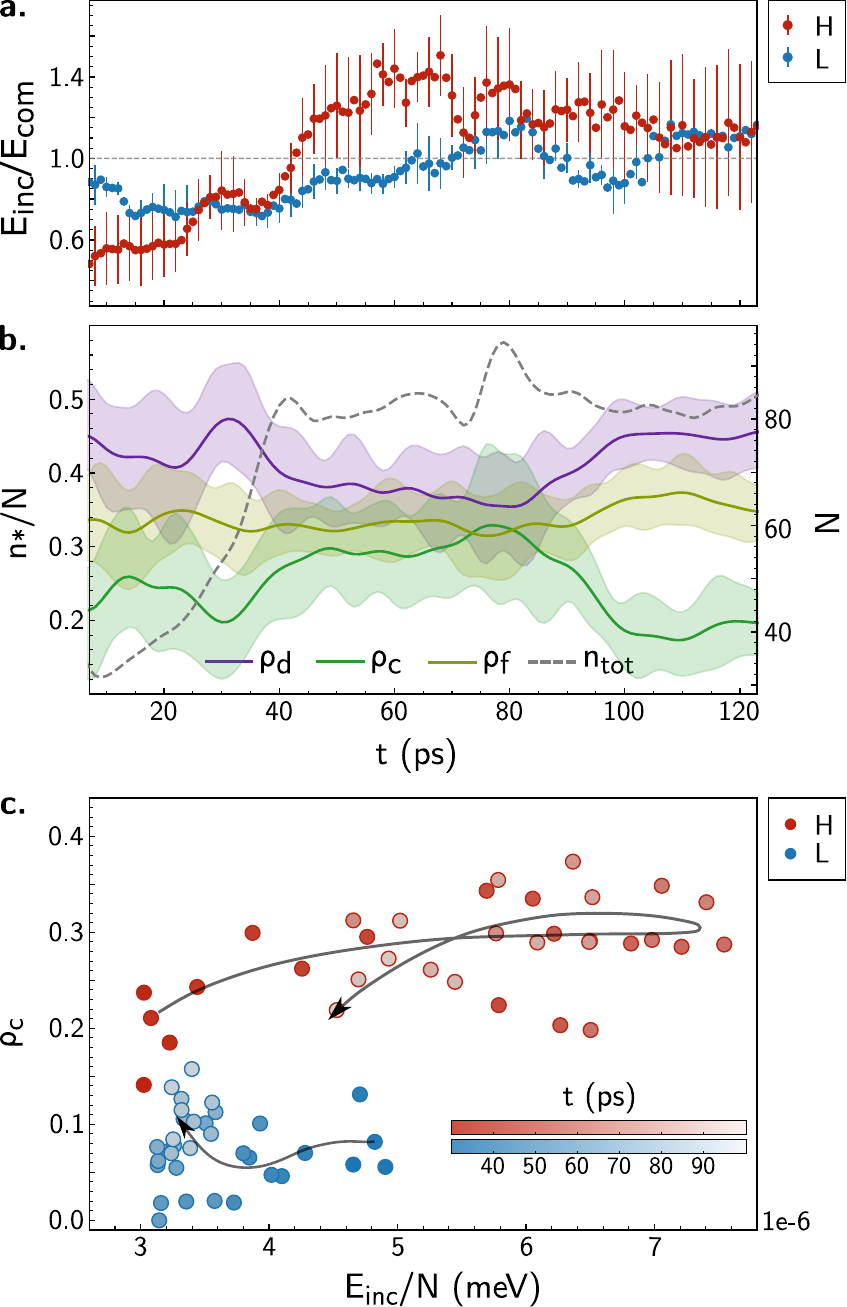}
	\caption{\textbf{a.} Ratio between the incompressible and compressible kinetic energy for the two configurations.
	\textbf{b.} Fractions of dipoles ($\rho_d$, purple line), clusters ($\rho_c$, green line) and free vortices ($\rho_f$, yellow line) for the H configuration. The total number of vortices in time is shown by the dashed gray line.
	\textbf{c.} Relative number of clusters versus mean incompressible energy per vortex. Arrows represent the direction of time.}
	\label{fig:3}
\end{figure}

\paragraph*{Temporal evolution and onset of clustering. -}
The vortex tracking allows us to compute the correlation function $C = \frac{1}{N}\sum_{i=1}^{N}c_i$ for each time frame, with $N$ the total vortex number and $c_i=1$ if the circulation of the nearest neighbor of the i-th vortex has the same sign, or $c_i=-1$ if it has the opposite sign. 
Increasing values of C correspond to higher energetic states of the vortex gas.
In the Onsager model, negative temperatures are associated to $C>0$, whereas $C=0$ corresponds to the infinite temperature limit (maximum entropy) and $C=-1$ is the lowest (positive) temperature~\cite{kraichnan1967inertial,simula2014prl,Groszek2018}. 
Typically, during the spontaneous evolution of a polariton fluid, $C$ is negative and tends towards the lowest energy state, $C=-1$, due to the dissipative nature of the system and the spontaneous nucleation of dipoles (additional experimental data in \cite{SM}). In Fig.~\ref{fig:2}c, we show instead that $C$ increases between $40$~\si{\pico\second} and $80$~\si{\pico\second}.
This shows that, despite the finite polariton lifetime ($\approx\SI{100}{\pico\second}$), the vortex gas evolves towards more energetic configurations.
After $80$~\si{\pico\second}, eventually dissipation prevails and correlation starts decreasing. This behavior is visible only in H (red points in Fig.~\ref{fig:2}c), while in L (blue points) the correlation remains almost constant $C\approx-0.7$ during the whole dynamics.

The observed increase of $C$ requires the injection of incompressible kinetic energy into the system. In a closed system such as an atomic BEC, even in the absence of a constant energy injection, this is explained by means of a evaporative-heating mechanism~\cite{simula2014prl}. The vortex gas undergoes vortex-pairs annihilation while conserving the total energy of the system, leading to an increase of the mean energy per vortex: the number of vortices decreases with time, and the few remaining vortices tend to form small clusters~\cite{johnstone2019evolution, Kanai2021}. In our open system, the simultaneous presence of sound waves and vortices enables additional mechanisms. In the following we show that the kinetic energy of sound waves is efficiently transformed into kinetic energy of the vortex gas.

In Fig.~\ref{fig:3}a, the ratio of the compressible and incompressible kinetic energy is shown for H and L as a function of time. In H (red points), the compressible component is transformed into incompressible kinetic energy, which becomes dominant starting from 40~\si{\pico\second}. On the contrary, in L (blue points), the sound waves component is higher than the incompressible one during the whole dynamics, becoming roughly the same at later times. In Fig.~\ref{fig:3}b, the three vortex species fractions, namely dipoles ($\rho_d$), clusters ($\rho_c$) and free vortices ($\rho_f$) are shown for the H configuration, along with the total number of vortices (dotted line). While in the first $40$~\si{\pico\second} the increase of the incompressible kinetic energy occurs through the nucleation of new vortices, the further increase up to $80$~\si{\pico\second} occurs with a constant number of total vortices. In this interval of time, the increase of the number of clusters occurs at the expenses of dipoles, showing that the additional incompressible kinetic energy forces the vortex gas to form small clusters.
This is even clearer in Fig.~\ref{fig:3}c, where we plot the clustered fraction of vortices versus the average incompressible kinetic energy \textit{per vortex}, showing the opposite dynamical evolution of H and L configurations (see \cite{SM}).

\paragraph*{Energy spectrum. -}
\begin{figure}
	\centering
    \includegraphics[width=\columnwidth]{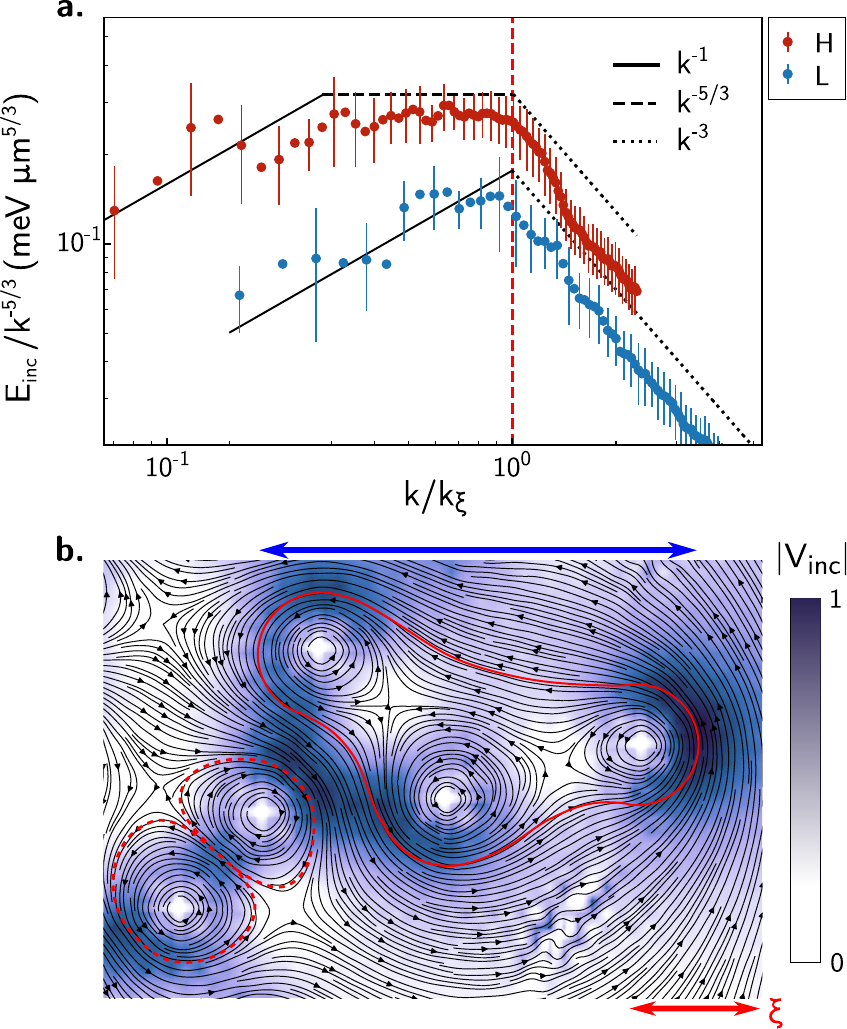}
	\caption{\textbf{a.} Energy spectrum for H (red) and L (blue), integrated in the time interval between 60~\si{\pico\second} and 80~\si{\pico\second} and  normalized to unity integral energy. The k-axis is rescaled to account for the different healing length (represented by a red dashed line) in H and L (see Methods). \textbf{b.} The incompressible velocity field around a configuration made of a dipole (dashed red line) and a cluster of three vortices with the same sign (solid red line). The background heat map represents the modulus of the incompressible velocity. To enhance the visibility, the region close to the vortex core is filtered out and the map is saturated at $|V_{inc}|=1.6$~\si{\micro\meter\per\pico\second}. The ``size'' of the cluster, indicated by the blue line, is $\sim3.4~\xi$.}
	\label{fig:4}
\end{figure}
Fig.~\ref{fig:4}a shows the incompressible kinetic energy spectrum measured in H and L configurations between 60~\si{\pico\second} and 80~\si{\pico\second}.
The appearance of a Kolmogorov-like $k^{-5/3}$ scaling law (horizontal, dashed-black line), associated to the inverse energy cascade~\cite{KOLMOGOROV1941,kraichnan1967inertial}, is clearly visible in H at wavevectors smaller than $k_\xi$, which corresponds to the injection scale of vortices (inverse healing length). Despite the small range of scale involved, this is the first direct measurement of an energy spectrum showing the inverse cascade in a 2D quantum fluid.

The separation of scales at the wave vector $k_{\xi}$ appears for both H and L. In the ultraviolet range $k>k_{\xi}$, both H and L show a $k^{-3}$ decay (dotted black line), the expected scaling law associated to the internal structure of a quantum vortex~\cite{Bradley2012}. In L, the infrared spectrum ($k<k_{\xi}$) tends towards $k^{-1}$ (solid black line), which can be derived from the velocity distribution of a collection of vortices in the far-field~\cite{Bradley2012}. On the contrary, in H, the $k^{-5/3}$ spectrum extends from $k_{c}<k<k_{\xi}$, with $k_c$ that approximately corresponds to the inverse of the typical spatial size of the clusters, $l_c\simeq (3\textrm{-}4)\xi$. In Fig.~\ref{fig:4}b, we show the incompressible velocity field, as extracted directly from the experiment in H, around a dipole and a cluster of three vortices. In the cluster, the velocity field is arranged on a larger spatial scale, with the flow circulating externally to the three vortices. Dimensional analysis shows that the time required to form clusters of that size is of about 20~\si{\pico\second} in our system~\cite{SM}.

To confirm our findings, we perform simulations of the Gross-Pitaevskii equation for the polariton field~\cite{Comaron2021}.
The appearance of the $k^{-5/3}$ scaling law in the incompressible kinetic energy spectrum is observed as well in numerical simulations and corresponds to an increase of both the correlation function and the incompressible kinetic energy per vortex~\cite{SM}.
As observed in both experiment and simulations, the dynamics is faster when the polariton density is larger and the intervortex distance is comparable to the healing length, resulting in an effective increase of the interactions between vortices. 

In conclusion, in this work we demonstrate the possibility of exploring turbulent states in quantum fluids of light. These results show the first evidence of the inverse energy cascade in dissipative quantum fluids, along with the onset of vortex clustering on timescales of few tens of picoseconds.
Importantly, we can decouple the compressible and incompressible contributions to the kinetic energy, showing that the energy required to start the clusterization dynamics is provided to the vortex gas by the dissipation of sound waves.
Finally, we show that the optical measurement of the velocity field allows an unprecedented control over the dynamics of the vortex gas, enabling a new series of experiments to be performed in nonlinear optical systems such as semiconductor microcavities, nonlinear crystals and laser beams coupled to hot atomic vapours or Rydberg atomic states~\cite{Claire2020, Guohai2020, Piekarski2021, Clark2020}.

\

\paragraph*{Methods. -}
\subparagraph*{Experiment}: The ring potential (diameter $R\simeq150$~\si{\micro\meter}) is realized by shaping a continuous-wave (CW) laser beam ($\lambda =735$~\si{\nano\meter}) with a spatial light modulator. The local energy shift of the polariton resonance, due to the high exciton density induced by the CW pump, is able to confine the polariton superfluid within the potential barrier (Fig.~\ref{fig:1}a). 
The polariton superfluid is quasi-resonantly injected ($\lambda\approx 773$~\si{\nano\meter}) by focusing a pulsed beam (pulse length~$\sim2$~\si{\pico\second}) in a Gaussian spot with a beam radius $w\simeq17$~\si{\micro\meter} at the center of the ring potential (Fig.~\ref{fig:1}b). Given the repetition rate of the pulsed pump (80 MHz) and the typical integration time of 1 ms, each time frame is obtained from the integration of a large number of pulses, giving as a result the average vortex distribution mediated over a large number of events. Additionally, our analysis is the result of an average of four measurements for each case, obtained by translating the sample in plane to exclude effects due to its morphology. The sample used is a planar Al$_x$Ga$_{1-x}$As microcavity with aluminium fractions of 0.2 and 0.95 in the distributed Bragg reflectors and 12 quantum wells of GaAs embedded in the cavity layer. The sample is kept at a cryogenic temperature of $\sim 5$ K. The measurements are taken in reflection configuration, with counter-polarized detection with respect to the polarization of the exciting laser. This is possible due to the precession of the polarization typically observed in polariton microcavity as a consequence of TE-TM splitting~\cite{kavokin2004}.

From the separation of scales observed in the energy spectrum (Fig.~\ref{fig:4}a) we are able to estimate the healing length in the $60$--$80$~\si{\pico\second} interval to be $\xi\simeq 4.6$~\si{\micro\meter} in H and $\xi\simeq 10.6$~\si{\micro\meter} in L.

In the energy decomposition, the density of the system is normalised at each time frame so that $\int\rho(\mathbf{r})\mathrm{d}\mathbf{r}=1$, to rule out non-relevant physical behaviours stemming from the modulation of the density throughout the evolution due to the measurement technique employed. 

In Fig.~\ref{fig:4}a, the energy spectrum is displayed for a range of wavevectors that spans from the inverse diameter of the trap $k_{t}=2\pi/R$ to $k_{r}=2\pi/r$, with $r=2$~\si{\micro\meter} (roughly the optical resolution of our optical setup), both rescaled by the respective $k_{\xi}=2\pi/\xi$ for H and L.

\

\subparagraph*{Simulations:}
To further confirm our findings, we perform simulations of the equation of motions for the polariton field $\psi=\psi(\textbf{r},t)$~\cite{carusotto2013quantum,Comaron2021} within the Truncated Wigner formulation~\cite{carusotto2013quantum} which reads ($\hbar=1$):
\begin{equation}
\hspace{-4mm}i d \psi_c = dt \bigg[ - \frac{
	\nabla^2}{2 m } +
g|{\psi_c}|^2_{-} 
+ \frac{i}{2} \gamma_{c} + V(\textbf{r}) \\ \bigg]
\psi_c +  dW_c
\label{eq:SGPE_pol}
\end{equation}
where $m$ is the polariton mass, {$g$ is} the polariton-polariton interaction strength, $\gamma_{c} = 1/\tau_c$ is the polariton loss rate, corresponding to the inverse of the polariton lifetime $\tau_c$, $dW$ the stochastic Wiener noise and $|{\psi}|^2_{-}$ the renormalized density $|{\psi}|^2_{-} \equiv \left(\left|{\psi} \right|^2 - {1}/{{(2dV)}} \right)$ which {includes} the subtraction of the Wigner commutator contribution (here $dV=a^2$ corresponds to the {volume element} of {our $2d$ grid of} spacing $a$). The zero-mean white Wiener noise $dW$ fulfils $\left <dW(\textbf{r},t)dW(\textbf{r}^\prime,t)\right>~=~0, \left < d W^*(\textbf{r},t)dW(\textbf{r}^\prime,t)\right> = \gamma_c/2 \delta_{\textbf{r},\textbf{r}^\prime} dt$.
Parameters are chosen in order to correspond with the microcavity used in the experiments, and reads $m = 0.22$~\si{\pico\second\milli\electronvolt\per\micro\meter\squared} $= 3.52 10^{-35}$~\si{\kilo\gram}, $g = 5\times 10^{-3}$~\si{\milli\electronvolt\micro\meter\squared}. The fluid of light is confined in a hard-bounded annular potential $V(\textbf{r})$ with a radius $r = 70$~\si{\micro\meter}.

The dynamics is initiated by a central Gaussian profile which provides the initial expansion, and whose local phase, being constant throughout space, accounts for the quasi-resonant nature of the initial experimental impulse.
The increase of the detuning of the pump in the experimental case is controlled by the intensity of the initial profile, which also corresponds to an increment of the density of particles, as well as the velocity of the outward particle flux.
Simulations for different blue-shifts are found to be in good agreement with the experimental results as shown in Figs.~\ref{fig:3}, \ref{fig:4}, and Figs.~S5, S6 in the Supplementary Materials. 
In Fig.~\ref{fig:3}b-c and Fig.~S6d-e (\cite{SM}), the fraction of vortices belonging to clusters is shown in experiments and simulations, respectively.
Increasing the injection energy, the formation of clusters is faster and occurs with higher probability. %
The faster dynamics is driven by stronger interactions between vortices, which are effectively increased at lower intervortex distances (i.e. when the total vortex density is increased), as observed in both experiments and simulations (Fig.~\ref{fig:3}b and Fig.~S6a in \cite{SM}). The relation between clusterization and vortex density is confirmed by additional analysis reported in the supplementary text.

\bibliography{biblio}

\end{document}


\title{Supplementary Material for: Onset of vortex clustering and inverse energy cascade in dissipative quantum fluids}

\author{R. Panico}
\affiliation{CNR NANOTEC, Institute of Nanotechnology, Via Monteroni, 73100 Lecce, Italy}
\affiliation{Dipartimento di Matematica e Fisica E.~De Giorgi, Universit\`a del Salento, Campus Ecotekne, via Monteroni, Lecce 73100, Italy}

\author{P. Comaron}
\affiliation{Institute of Physics, Polish Academy of Sciences, Al. Lotników 32/46, 02-668 Warsaw, Poland}
\affiliation{Department of Physics and Astronomy, University College London,
Gower Street, London, WC1E 6BT, United Kingdom}

\author{M. Matuszewski}
\affiliation{Institute of Physics, Polish Academy of Sciences, Al. Lotników 32/46, 02-668 Warsaw, Poland}

\author{A. S. Lanotte}
\affiliation{CNR NANOTEC, Institute of Nanotechnology, Via Monteroni, 73100 Lecce, Italy}
\affiliation{INFN, Sez. Lecce, 73100 Lecce, Italy}

\author{D. Trypogeorgos}
\affiliation{CNR NANOTEC, Institute of Nanotechnology, Via Monteroni, 73100 Lecce, Italy}

\author{G. Gigli}
\affiliation{CNR NANOTEC, Institute of Nanotechnology, Via Monteroni, 73100 Lecce, Italy}
\affiliation{Dipartimento di Matematica e Fisica E.~De Giorgi, Universit\`a del Salento, Campus Ecotekne, via Monteroni, Lecce 73100, Italy}

\author{M. De Giorgi}
\affiliation{CNR NANOTEC, Institute of Nanotechnology, Via Monteroni, 73100 Lecce, Italy}

\author{V. Ardizzone}
\affiliation{CNR NANOTEC, Institute of Nanotechnology, Via Monteroni, 73100 Lecce, Italy}
\affiliation{Dipartimento di Matematica e Fisica E.~De Giorgi, Universit\`a del Salento, Campus Ecotekne, via Monteroni, Lecce 73100, Italy}

\author{D. Sanvitto}
\thanks{Corresponding author: daniele.sanvitto@cnr.it}
\affiliation{CNR NANOTEC, Institute of Nanotechnology, Via Monteroni, 73100 Lecce, Italy}

\author{D. Ballarini}
\affiliation{CNR NANOTEC, Institute of Nanotechnology, Via Monteroni, 73100 Lecce, Italy}
\affiliation{Dipartimento di Matematica e Fisica E.~De Giorgi, Universit\`a del Salento, Campus Ecotekne, via Monteroni, Lecce 73100, Italy}

\begin{abstract}
\end{abstract}

\maketitle

\begin{figure}
\centering
\includegraphics[width=\columnwidth]{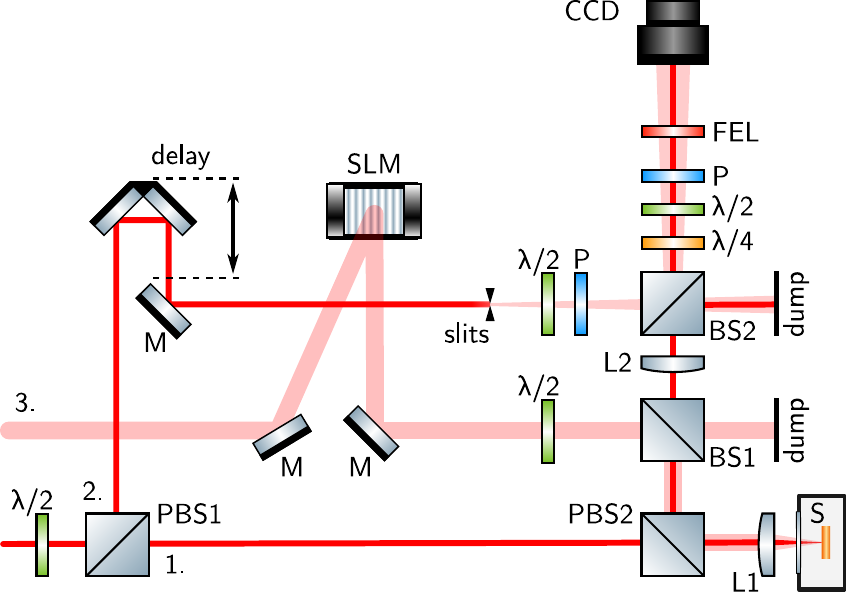}
    \caption{A schematic representation of the experiment. The pulsed laser beam is split in two (1., 2.) by a polarizing beam splitter (PBS1), coupled with a half-waveplate to tune the power in each arm. The first beam (1.) arrives focused on the sample (S), with a linear polarization, through a 5~\si{\centi\meter} camera objective (L1). The photoluminescence from the sample is collected by a second polarizing beam splitter (PBS2) which also serves as a first filter for the laser beam. The second beam (2.) hits a retroreflector mounted on a moving stage, which allows for fine adjustments of its optical path length. This beam then passes through an iris to create a spherical light source so that, after its propagation, we get a reference beam with a relatively flat phase. The two beams (1., 2.) are then put back together at a beam splitter (BS2). A half-waveplate and a polarizer ensure that the reference has the same polarization as the emission. A Bessel phase is imprinted on a CW laser (3.) with a spatial light modulator (SLM); this beam enters the sample from a beam splitter (BS1) placed in the detection line and a half-waveplate is used to tune the power that passes through PBS2. The camera objective (L1) Fourier transforms this beam into a ring, that is then used to confine the superfluid. The image from the sample is reconstructed on the CCD with a 100~\si{\centi\meter} lens (L2), after passing through a set of waveplates and a polarizer to extinguish the resonant laser and a long-pass filter to block the non-resonant one. The beam sizes and the length of the lines are not to scale, and unnecessary reflections from the beam splitters (BS) have been omitted.}
	\label{fig:setup}
\end{figure}

\paragraph{Experimental setup. -}
The schematic representation of the experimental setup, with a complete description of the optics and the lines, is shown in Fig.~\ref{fig:setup}.
The pulsed laser (2~\si{\pico\second}, 80~\si{\mega\hertz}) is used to quasi-resonantly excite the system  (beam 1. in Fig.~\ref{fig:setup}) at $\sim772.9$~\si{\nano\meter} for the H case and at $\sim773.4$~\si{\nano\meter} for L. The trapping, as discussed in the main text, is made with an out of resonance CW laser at $\sim730$~\si{\nano\meter}. To shape the beam as a ring we employed a phase-only SLM with a Bessel pattern, working with a diffraction grating to carry on only the modulated part of the beam.

The signal and the reference beam form an angle (not visible in Fig.~\ref{fig:setup}), for an easier separation of the information contained in the interferograms that we measure.

\vspace{0.5cm}

\begin{figure}
\centering
\includegraphics[width=\columnwidth]{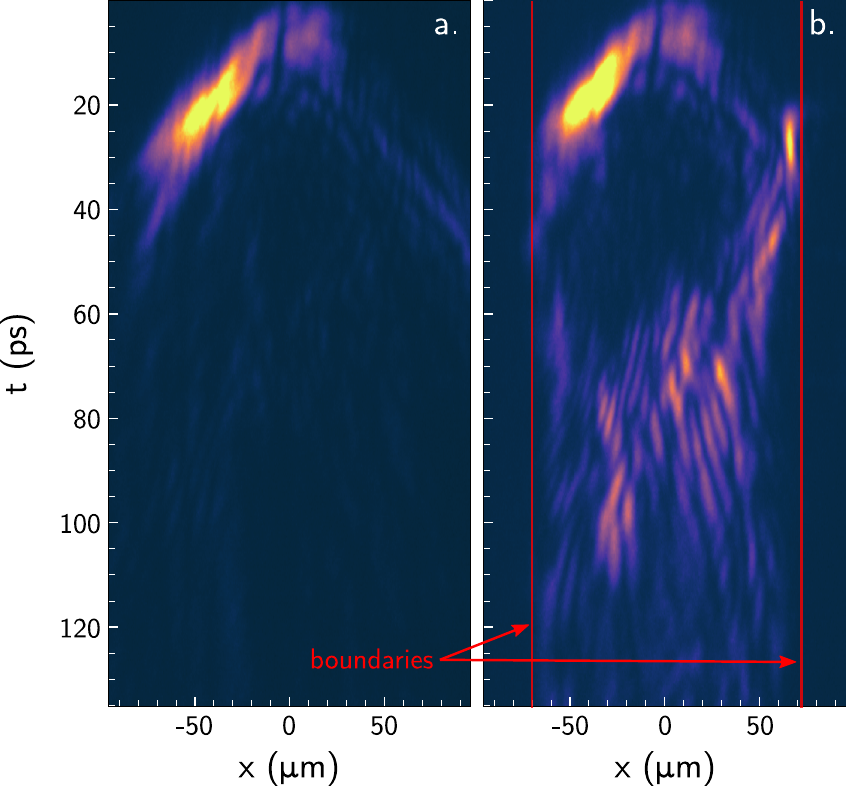}
    \caption{Streak camera images of the central slice of the superfluid without (a) and with (b) the optical confinement. The red vertical lines represent the size of the barrier.}
	\label{fig:streak}
\end{figure}

\paragraph{Optical confinement of the superfluid. -}
The prove the effectiveness of the optical confinement by means of an out of resonance ring, we looked at the time evolution of the superfluid with the aid of a streak camera. Figure~\ref{fig:streak} shows the evolution of a central horizontal slice of the condensate without (left) and with boundaries (right). In the latter the dynamics of the system can be followed for a much longer time, whereas without any type of confinement the superfluid quickly expands.

\vspace{0.5cm}

\paragraph*{Helmholtz decomposition of the velocity. -}
To disentangle the velocity into its incompressible and compressible
parts, we adopt the Helmholtz decomposition for vector
fields. Given $\boldsymbol{v}(\boldsymbol{x})$, we can write it as the
sum of a potential and a divergence-free contribution:
$\boldsymbol{v}(\boldsymbol{x})={\bf \nabla}\phi(\boldsymbol{x}) +
\nabla\times\boldsymbol{A}(\boldsymbol{x})$, where $\phi$ and
$\boldsymbol{A}$ are a scalar and a vector field, respectively.
In the Fourier space, the decomposition has a straight-forward
implementation. Indeed, given the vector field in real space
$\boldsymbol{v}(\boldsymbol{x})$:
\begin{equation*}
\boldsymbol{V}(\boldsymbol{k}) =  \frac{1}{2 \pi} \int e^{-i \boldsymbol{k} \cdot \boldsymbol{x}} \boldsymbol{v}(\boldsymbol{x}) d{\cal S}_x\,,
\end{equation*}
which can be decomposed as
\begin{equation}
\boldsymbol{V}(\boldsymbol{k}) = i\boldsymbol{k}V_{\phi}(\boldsymbol{k}) + i\boldsymbol{k} \times \boldsymbol{V}_{\boldsymbol{A}}(\boldsymbol{k})
\label{helmholtz}
\end{equation}
where we used the scalar and vector fields:
\begin{align*}
    &V_{\phi}(\boldsymbol{k}) = - i\,\frac{\boldsymbol{k}\cdot \boldsymbol{V}(\boldsymbol{k})}{||\boldsymbol{k}||^2}\\
    &\boldsymbol{V}_{\boldsymbol{A}}(\boldsymbol{k}) = i\,\frac{\boldsymbol{k}\times \boldsymbol{V}(\boldsymbol{k})}{||\boldsymbol{k}||^2}.
\end{align*}

Once evaluated these in Fourier space, back to real space we finally have:
\begin{align*}
    \nabla\phi(\boldsymbol{x})&=\int e^{i \boldsymbol{k} \cdot \boldsymbol{x}}\,i \boldsymbol{k}V_{\phi}(\boldsymbol{k}) d{\cal S}_k\\
    \nabla\times\boldsymbol{A}(\boldsymbol{x})&=\int e^{i \boldsymbol{k} \cdot \boldsymbol{x}}\,i \boldsymbol{k} \times \boldsymbol{V}_{\boldsymbol{A}}(\boldsymbol{k}) d{\cal S}_k
\end{align*}
which are the potential and divergence-free components of the velocity field, respectively.\\

\vspace{0.5cm}

\paragraph*{Kinetic energy decomposition. -} At any given time of the temporal evolution, the total energy of the
quantum fluid can be written as the sum of hydrodynamic kinetic,
quantum pressure, potential, and interaction contributions~\cite{Barenghi_book,Bradley2012}:
\begin{equation}
  E_\mathrm{tot} = E_\mathrm{kin} + E_\mathrm{q} + E_\mathrm{V} + E_\mathrm{I} \,,
  \label{eq:Etot}
\end{equation}
where

\begin{eqnarray*}
E_\mathrm{kin}&=&m/2 \int n(\boldsymbol{x},t) |\boldsymbol{v}(\boldsymbol{x},t)|^2 d{\cal S}_x\,,\\
E_\mathrm{q}&=&\hbar^2/(2m) \int |\sqrt{n(\boldsymbol{x},t)}|^2 d{\cal S}_x\,,\\
E_\mathrm{V}&=&\int n(\boldsymbol{x},t) V(\boldsymbol{x},t) d{\cal S}_x\,,\\
E_\mathrm{I}&=& g/2 \int n(\boldsymbol{x},t)^2 d{\cal S}_x.\\
\end{eqnarray*}
Clearly in our case, this is not conserved in time, since we loose
polaritons.

The hydrodinamics kinetic (from now on, simply kinetic) energy
$E_\mathrm{kin}=m/2 \int n(\boldsymbol{x},t)
|\boldsymbol{v}(\boldsymbol{x},t)|^2 d{\cal S}_x$ can be further
decomposed in \textit{incompressible} and \textit{compressible}
components, which are attributed to the kinetic energy of quantum
vortices and of the sound excitation, respectively. Following
Refs.~\cite{Nore1997,Horng2009,Bradley2012} we define a
density-weighted velocity field $\boldsymbol{u}(\boldsymbol{x},t) =
\sqrt{n(\boldsymbol{x},t)} \boldsymbol{v}(\boldsymbol{x},t)$ and
decompose it as :
\begin{equation}
    \boldsymbol{u}(\boldsymbol{x},t) = \boldsymbol{u}^\mathrm{inc}(\boldsymbol{x},t) + \boldsymbol{u}^\mathrm{comp}(\boldsymbol{x},t),
\label{eq:u_inc_com}
\end{equation}
where the incompressible and compressible field satisfies
$\boldsymbol{\nabla} \cdot \boldsymbol{u}^\mathrm{inc} = 0$ and
$\boldsymbol{\nabla} \times \boldsymbol{u}^\mathrm{com} = 0$,
respectively. In Fourier space, the total incompressible kinetic
energy reads as
\begin{equation}
  E_\mathrm{kin}^\mathrm{inc} = \frac{m}{2} \sum_{i=x,y}\int  |U_i^\mathrm{inc}(\boldsymbol{k})|^2 d{\cal S}_k,
\label{eq:totkin}  
\end{equation}
with
\begin{equation}
  U_i^\mathrm{inc}(\boldsymbol{k}) = \frac{1}{2 \pi} \int e^{-i \boldsymbol{k} \cdot \boldsymbol{x}} u^\mathrm{inc}_i(\boldsymbol{x})\,d{\cal S}_x\,.
\label{eq:Uinc}  
\end{equation}
Finally, since the system geometry is isotropic we can simply consider
the longitudinal energy spectrum $E_\mathrm{kin}(k)$
(where we have dropped the incompressible label for simplicity), which
is obtained by integrating over the azimuthal angle as
\begin{equation}
    E_\mathrm{kin}(k) = \frac{m}{2} k \sum_{i=x,y} \int_\Omega d\Omega_k |U_i^\mathrm{inc}(\boldsymbol{k})|^2\,.
\label{eq:E_long}
\end{equation}

\vspace{0.5cm}

\paragraph{Turbulent estimates -} To find the signature of turbulent behaviour, the simplest observable
is a two-point statistical object. In the section above, we have
defined the isotropic (or longitudinal) energy spectrum
$E_\mathrm{kin}(k) = m \pi k \langle |{\bf
  U}^\mathrm{inc}(k)|^2\rangle$, from which the total incompressible
kinetic energy can be obtained as the sum over all $k=|{\bf k}|$,
$E_\mathrm{kin} = \int E_\mathrm{kin}(k) \,dk$.

Here we want to dimensionally estimate the characteristic time-scales of the turbulent inverse energy cascade~\cite{BEreview}, and see if these are compatible with our experimental measurements.
To do that, the starting point is the observation that $E_{\mathrm{kin}}(k)\sim k^{-5/3}$ in some range of wavenumbers, that in our experiments is $k_c < k < k_{\xi}$, where $k_c$ is the inverse of to the typical cluster scale $l_c$, and $k_{\xi}$ is the inverse of the healing length $\xi$. The characteristic times $\tau_{lc}$ of the clusters at scale $l_c\simeq 1/k_c$ in the turbulent inverse cascade obeys the scaling law $\tau_{lc} \simeq l_c^{2/3}$ (or equivalently the characteristic frequencies of cluster of size $k$ obey $\omega_k \simeq k^{2/3}$ for $k_c < k < k_{\xi}$).

To make the scaling argument dimensionally correct, we write the characteristic time as: $\tau_{l}= \left[l/(\delta_l u)\right]$, where $\delta_l u$ is the characteristic velocity fluctuation at scale $l$, and an $O(1)$ adimensional constant is absorbed into the characteristic velocity value.

Since the Kolmogorov scaling is observed in the region of wavenumbers $k_c < k < k_{\xi}$, it exists a corresponding region in real space $\xi< l < l_c$ such that: $$\frac{\delta_l u^3}{l} = cost\,$$ which extends down to the healing length. Hence we obtain: $\delta_l u \simeq (\delta_{\xi} u) (l/\xi)^{1/3}$, for all $\xi <l <l_c$.

Then we can use the experimental values for the typical velocity at the healing length, and the healing length itself: in the experiment H, these are $\delta_{\xi} u = \xi/ \tau_{\xi}= (0.7 \pm 0.1)$~\si{\micro\meter\per\ps}, where $\xi=(4.6 \pm 0.5)$~\si{\micro\meter}, and $\tau_{\xi}=(6.6 \pm 0.7)$~\si{ps}.

By applying the $\tau_{l}$ estimate to the scale of the clusters $l_c  \simeq 4 \xi$, we get: $\tau_{l_c}= \left[l_c/\delta_{l_c} u\right]\simeq (16.5 \pm 1.5)$~\si{ps}, which tells us that the characteristic time of the clusters at the observed scale $l_c$ is comparable to that observed in the experiments.

\vspace{0.5cm}

\paragraph{Previous configurations. -}
\begin{figure}
\centering
\includegraphics[width=\columnwidth]{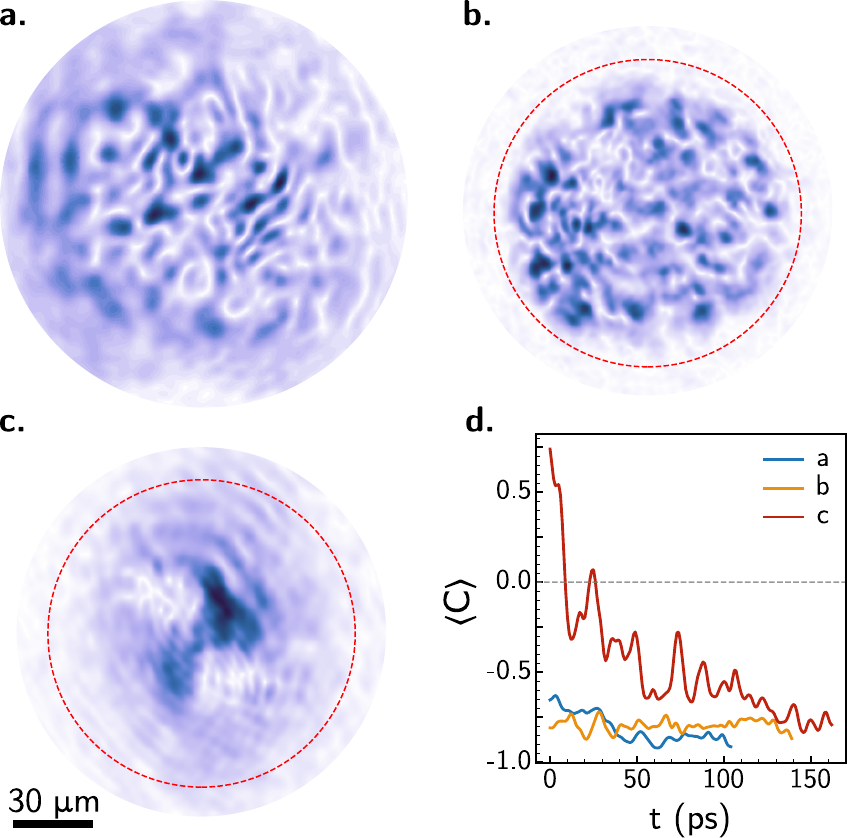}
\caption{Snapshots of the configurations used before the successful
  one described in the paper. (\textbf{a}) Without a trapping
  potential the superfluid expands way faster than the average speed
  of vortices, causing an overall increase of the mean
  distances. (\textbf{b}) Confining the superfluid in a circular
  potential and increasing the polariton density stabilised the number
  of vortices in time but nucleation (and annihilation) is still the
  predominant process. (\textbf{c}) Starting from two clusters the
  superfluid quickly becomes inhomogeneous and dipoles start to
  appear. (\textbf{d}) Value of the first order correlation function
  for the configurations in a, b, and c.  }
	\label{fig:old}
\end{figure}
Several configurations were tested before the one used in the main text, all of which employed a second SLM to imprint in a controllable any desired vortex distribution as initial configuration on a flat polariton superfluid.

Free expansion (Fig.~\ref{fig:old}a): without the trapping potential, polaritons are free to expand indefinitely and the inter-vortex distance increases with time. In this configuration, vortex-vortex interactions are too weak and the correlation function C decreases from C=-0.75 at t=0 towards C=-1 at later times.

High vortex density imprinted externally (Fig.~\ref{fig:old}b).: to increase the vortex-vortex interaction, a random distribution of a large number of vortices and antivortices with C=0 can be imprinted in the pulsed beam directly by the SLM. In this case, the condensate fills the whole trap since the beginning, without initial expansion. However, using this technique to inject a random distribution of vortices in a homogeneous polariton fluid is not effective: even if the number of vortices is more or less constant over time, we observe that the value of the first order correlation function is negative from the very beginning of the polariton dynamics. Indeed, vortices and antivortices just annihilate before reaching the sample during propagation in the linear medium (air), creating a very inhomogeneous condensate with a large number of dipoles and C decreasing quickly with time towards C=-1.

Externally imprinted clusters (Fig.~\ref{fig:old}c): we inject a flat polariton fluid with two clusters of vortices with positive and negative circulation, respectively, corresponding to the configuration with C=1. Also in this case, the polariton fluid is injected with the same size as the trap, without initial expansion. We observe the breaking of the cluster structure in time, accompanied with the nucleation of a large number of dipoles in the low-density regions of the condensate, quickly bringing the correlation towards C=-1.

Fig.~\ref{fig:old}d shows the temporal evolution of the first order correlation function in the three cases, where C is always decreasing towards C=-1. Indeed, we find that the collision of the expanding polariton fluid against the potential barrier is the best strategy to induce the hydrodynamic nucleation of a large number of vortices and antivortices in this system, avoiding the annihilation of vortices of opposite sign that occurs when external imprinting techniques and propagation through linear materials are used. Moreover, the formation of sound waves enables the injection of incompressible kinetic energy into the vortex distribution, as shown in Fig.~3a of the main text.

\vspace{0.5cm}

\paragraph{Low energy case vortex distribution. -}
\begin{figure}
\centering
\includegraphics[width=\columnwidth]{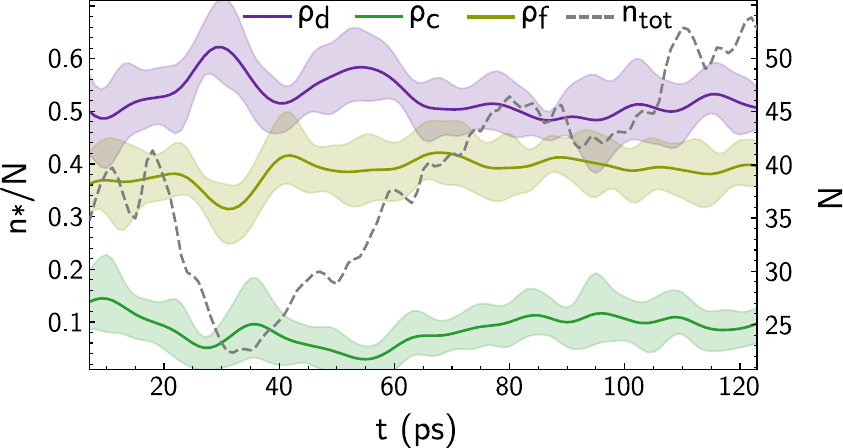}
    \caption{Fractions of dipoles ($\rho_d$, purple line), clusters ($\rho_c$, green line) and free vortices ($\rho_f$, yellow line) for the low energy configuration. The total number of vortices in time is shown by the dashed gray line.}
	\label{fig:ratios}
\end{figure}
We report in Fig.~\ref{fig:ratios} the three vortex species fractions, namely dipoles ($\rho_d$), clusters ($\rho_c$) and free vortices ($\rho_f$), together with the total number of vortices, for the L case. Contrary to what is show in Fig.~3b of the main text we can observe in this case slower dynamics of the superfluid, with the formation of the vortices due to the backflow taking place at a later time ($\sim 10$~\si{\pico\second}). The three populations remain approximately constant throughout the evolution, with the dipoles being the highest fraction at all times.

\vspace{0.5cm}

\begin{figure}
\centering
\includegraphics[width=\columnwidth]{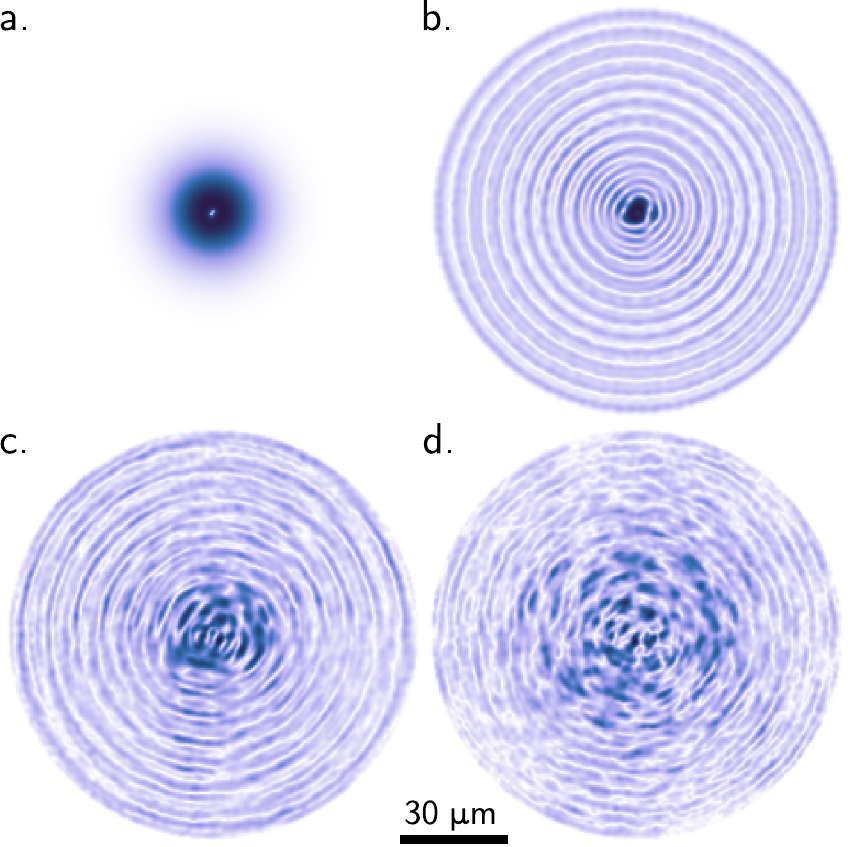}
    \caption{Temporal evolution of the simulated polariton fluid: the density field within the trap is shown at the starting of the evolution, (a) $t=0$ when the wavefunction has a Gaussian profile. At intermediate times {$t=70~\si{\pico\second}$} (b) and {$t=120~\si{\pico\second}$} (c) the wave-turbulence generates topological defects that at later time {$t=190~\si{\pico\second}$} (d) proliferate throughout the whole sample.}
\label{fig:1_SM}
\end{figure}

\paragraph*{Numerical modelling and simulation results. ---}
The collective dynamics of the polariton fluid is described by means
of Eq.~\red{(1)} of the main text, which we solve using a eighth-order Runge-Kutta method on $N^2=512^2$ points numerical grid with discretization $ a \simeq 0.6~\si{\micro\meter}$. 
The turbulent dynamics of the quantum fluid is simulated by starting from a wavefunction with a Gaussian profile (and uniform phase) $\psi({\bf x},t=0) = A \exp(-(x^2 + y^2)/2 \sigma^2)$ (see Fig.~\ref{fig:1_SM}a), resembling the resonantly-pulsed excitation applied in the experiments. The parameter $A=\chi n_0^{1/2}$ controls the density of particles (which along the time evolution is in the order of $|\psi|^2 \sim 10^2$~\si{\per\micro\meter\squared}), as well as the velocity of the outward particle flux. 
Therefore, varying the amount of blue-shift in the system corresponds to tuning the parameter $\chi$.
For our numerical simulations, we choose the width of the initial Gaussian profile $\sigma = 24.5$~\si{\micro\meter}, so that the system average healing length is $\xi \sim 2$~\si{\micro\meter} for $\chi=0.7$ ($n_0$ is fixed to 1). 
A conservative dynamics ($\gamma = 0$) is adopted since it has been already successfully employed to simulate quasi-resonant pumping exciton-polaritons~\cite{Panico2020}~\red{[..]}. 
We are careful to simulate the system dynamics in a numerical grid which is at least double the size of the ring diameter, so to avoid artificial boundary effects.

An example of the temporal dynamics from numerical simulations is illustrated by the snapshots in Fig.~\ref{fig:1_SM}, showing the trapped condensate for $\chi = 0.7$.
During the first stage of the dynamics, the condensate radiates isotropically until it hits the hardly-bound confinement, eventually creating density fringes which emerge due to the interference with the reflected wave, Fig.~\ref{fig:1_SM}b.
The wave turbulence is then responsible for the generation a large number of vortices  which start to emerge from the centre of the trap (Fig.~\ref{fig:1_SM}c) and eventually expand throughout the whole sample, where they are free to proliferate (Fig.~\ref{fig:1_SM}d).
The single topological defects is numerically identified computing
phase gradients around closed paths of each grid point; more specifically when the circulation around a close path $C$ of double the size of the vortex healing
length is found to be approximately $2 \pi$ \cite{Comaron2021}.

\begin{figure}
	\centering
    \includegraphics[width=\columnwidth]{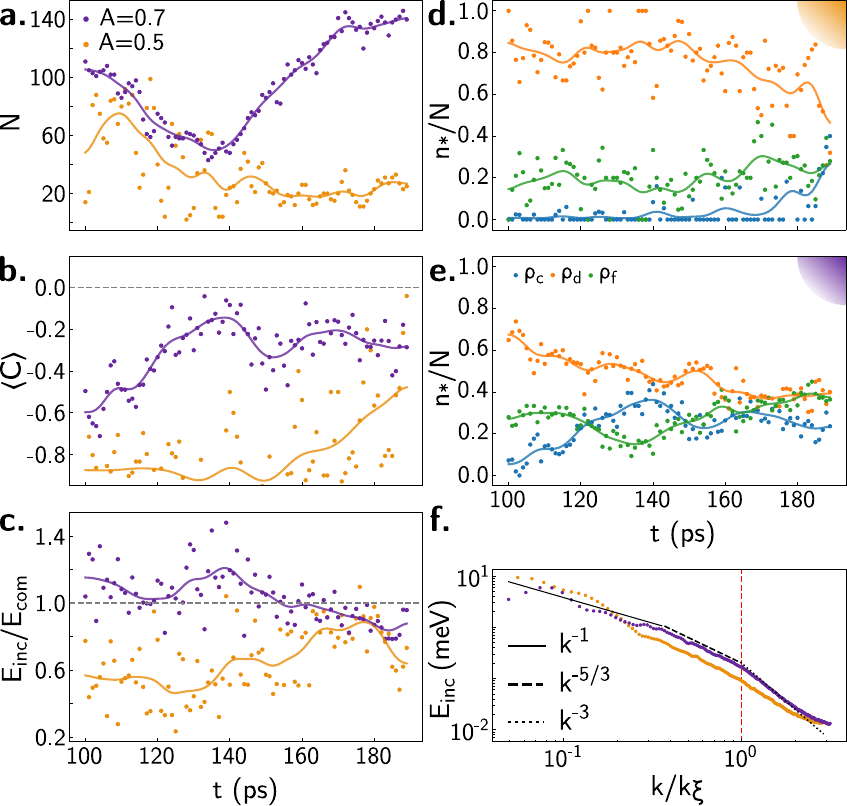}
	\caption{Physical observables of the system as extracted from the simulations for the lowest and highest detuning value. \textbf{a.}~Total number of vortices. \textbf{b.}~Value of the clustering correlation function, which show the increasing trend over time. \textbf{c.}~Ratio of the incompressible and compressible kinetic energy during the evolution. \textbf{d.-e.} Ratios of the number of vortices belonging to clusters ($\rho_c$), dipoles ($\rho_d$) or neither of them (free, $\rho_f$) for the low (d) and high (e) density case, respectively. \textbf{f.}~Incompressible kinetic energy spectra showing the Kolmogorov scaling in the time range $t \in [140-190]$~\si{ps}, for wavenumbers smaller than the inverse of the healing length and larger than those dominated by the far-field behaviour.}
	\label{fig:comparison}
\end{figure}
We proceed by investigating the turbulent properties of the fluid at different detuning rates.  We explored the dynamics by varying the detunings as $\chi=[0.5; 0.55; 0.6; 0.7]$.
Figure~\ref{fig:comparison} shows the physical observables calculated throughout the temporal evolution for L, the lowest detuning (yellow case), and H, the highest detuning (purple case).
The quantities are calculated after $100 \si{ps}$, when the vortices have been generated by the initial pulse.
In qualitative agreement with the experimental curves reported in the main script (see Fig.~2c and Fig.~3a,b), the numerical curves exhibit an enhanced clustering correlation function (Fig.~\ref{fig:comparison}b) for higher detunings. This fact is corroborated by the 
behaviour of the temporal ratio between incompressible and compressible energies (Fig.~\ref{fig:comparison}c) and the fraction between free, dipole and clustered vortices, reported in Fig.~\ref{fig:comparison}d and Fig.~\ref{fig:comparison}e, respectively.
We attribute this feature to stronger interactions between vortices, driven by the mean inter-vortex distance which decreases at higher defect densities. 
Noteworthy, it is in the late-time dynamics, between $140$ and $190$~\si{ps}, that the system presents the highest number of clustered vortices over dipoles and free vortices, correspondingly to an increment of incompressible energy per vortex and highest correlation function values.
Finally, adopting the methods explained in the following section, we compute the incompressible energy spectra for the different cases investigated.
Fig.~\ref{fig:comparison}f depicts the low- and high-detuned time-averaged spectra calculated in the late-time dynamics of the system. 
In qualitative agreement with the experimental results shown in Fig.~\red{4}a of the main text, the total amount of incompressible energy is found to be larger in the H case when compared to lower detunings. 
Importantly, the latter case also exhibit the  expected scalings for a turbulent quantum fluid~\cite{Bradley2012}, in the ranges of wavelength smaller than the inverse of $\xi$, as well as in the infra-red and ultra-violets part of the spectrum, as discussed in the main paper.

We conclude noting that exclusion of the stochastic term from the numerical integration of Eq.~\red{(1)} of the main text confirms that such a turbulent behavior does not arise from the spontaneous formation of vortices due to quantum fluctuations, as seen in previous works~\cite{comaron2018dynamical} \red{[..]}, but instead takes place at a mean-field level due to the strong wave turbulence generated by the presence of the barrier.

\vspace{0.5cm}


\bibliography{biblio_sim.bib}